\input harvmac
\input epsf

\Title{
\rightline{hep-th/0211038}}
{\vbox{\centerline{On DBI action of the non-maximally symmetric D-branes }
\centerline{ on SU(2) }}}

\medskip
\centerline{Gor Sarkissian\foot{e-mail: gor@ictp.trieste.it}}
\bigskip
\smallskip
\centerline{The Abdus Salam International Centre for Theoretical Physics}
\centerline{Strada Costiera 11, Trieste 34014, Italy}

\smallskip

\bigskip\bigskip\bigskip
\noindent
In this paper we study the non-maximally
symmetric D-branes on the SU(2) group 
discussed in a previous article (hep-th/0205097).
Using the two-form defined in hep-th/0205097 the DBI
action on the branes is constructed. 
This action is checked for its agreement with
CFT predictions.
The geometry of the branes is analyzed in detail, and the
singularities of branes covering the entire group are found. 
\vfill

\Date{11/02}

\lref\klsev{C. Klimcik and P. Severa, 
Open strings and D-branes in WZNW models,
Nucl. Phys. B488 (1997) 653, hep-th/9609112.}%

\lref\alschom{A. Alekseev and V. Schomerus, 
D-branes in the WZW model, Phys. Rev. D60, 061901 (1999),
hep-th/9812193.}%

\lref\gaw{K. Gawedzki, 
Conformal field theory: a case study, in Conformal Field Theory,
Frontier in Physics 102, Eds. Y. Nutku, C. Saclioglu, T. Turgut
( Perseus Publishing, 2000),
hep-th/9904145.}%
\lref\bdsnew{C. Bachas, M. Douglas and C. Schweigert, 
Flux stabilization of D-branes, JHEP 0005 (2000) 048,
hep-th/0003037.}%

\lref\malmosei{J. Maldacena, G. Moore and N. Seiberg, 
Geometrical interpretation of D-branes in gauged WZW models,
JHEP 0107 (2001) 046,
hep-th/0105038.}%

\lref\malmnat{J.\ Maldacena, G.\ Moore and N.\ Seiberg,
D-brane instantons and K-theory charges, JHEP 0110 (2001) 005,
 hep-th/0108100.}%

\lref\sark{G. Sarkissian, Non-maximally symmetric D-branes
on group manifold in the Lagrangian approach, JHEP 0207 (2002) 033,
hep-th/0205097.}

\lref\walzh{M. A. Walton and J.-G.  Zhou,
D-branes in asymmetrically gauged WZW models and axial-vector 
duality, hep-th/0205161.}

\lref\quella{ T. Quella, On the hierarchy of symmetry breaking D-branes 
in group manifolds, hep-th/0209157.}

\lref\arsch{A. Yu. Alekseev, A. Recknagel and V. Schomerus,
Non-commutative world-volume geometries: Branes on SU(2) and fuzzy
spheres, JHEP 09 (1999) 023, hep-th/9908040.}

\lref\afqsch{A. Yu. Alekseev, S. Fredenhagen, T. Quella
and V. Schomerus, Non-commutative gauge theory of twisted
D-branes, hep-th/0205123.}
\newsec{Introduction}

\lref\boris{P. Bordalo, S. Ribault and C. Schweigert, 
Flux stabilization in compact groups, JHEP 0110 (2001) 036,
hep-th/0108201.}
\lref\stanold{S. Stanciu, 
A note on D-branes in group manifolds: flux quantization and D0-charge,
JHEP 0010 (2000) 015, hep-th/0006145.}%
\lref\fkll{J. Fuchs, P. Kaste, W.Lerche, C. Lutken, C. Schweigert, J. Walcher,
Boundary Fixed Points, Enhanced Gauge Symmetry and Singular Bundles on
K3, Nucl. Phys. B598 (2001) 57-72, hep-th/0007145.}
\lref\figso{J.M. Figueroa-O'Farrill, S. Stanciu, D-brane charge, 
flux quantisation and relative (co)homology, JHEP 0101 (2001) 006,
hep-th/0008038.}
\lref\gawre{K. Gawedzki, N. Reis, WZW branes and gerbes,
hep-th/0205233.}

During the last years progress has been made in the understanding
of branes on group manifolds. In the pioneering work
of Klimcik and Severa \klsev\ additional
topological conditions required for writing down 
a well-defined WZW action on a world-sheet with boundary were formulated.
In that work also one-, two- and three-dimensional
branes on $SU(2)$ satisfying these conditions were found, but the symmetry
properties of those branes remained obscure.
These topological conditions were analyzed further 
in \figso\ , \gawre\ .

Following these findings, in \alschom\ maximally-symmetric
solutions to these conditions, which are 
quantized conjugacy classes, were found. These branes were
extensively studied and, by now, an almost complete
understanding of them has been established: 
their Lagrangian formulation \alschom\ , \gaw\ ,
the DBI action, stability properties \bdsnew\ ,
non-commutative geometry \arsch, \afqsch\ etc.

Recently, using T-duality and the boundary state formalism, in
\malmosei\ and \malmnat\ 
non-maximally symmetric branes on group manifolds
and parafermionic target spaces, called B-branes, were found, preserving
only part of the diagonal chiral algebra.
In \sark\ , the vectorial and axial symmetries, obtained with algebraic
methods in \malmnat\ , were checked and confirmed in the Lagrangian formulation.
It was also shown in \walzh\ that boundary conditions
defining B-branes on group manifolds admit axial gauging,
resulting in B-branes on parafermionic target spaces.
In \quella\ , a generalization of these methods was suggested,
associating symmetry breaking branes to certain quantized  chains 
of subgroup embeddings.

In this work we study the geometry and the DBI action of the 
non-maximally symmetric branes on SU(2) introduced in \malmosei\ ,
using the Lagrangian formulation suggested in \sark\ .   

In section 2 we analyze the geometry of the branes and find
that, generically, they are  $(U(1)\times S^2)/Z_2$.
For even values of $k$, branes covering the whole group,
in the complement of a big circle, exist and coincide
with one of the branes found in \klsev\ . 

In section 3, using the relation between the DBI action and the topological
two-form present in the Lagrangian formulation of the WZW
model on a world-sheet with boundary 
 \boris\ , \stanold\ ,
we compute the DBI action on these branes. We show that the target space
mass computation is in agreement with the CFT prediction.

In the appendix we present in detail the two-form computation.

\newsec{ Geometry of the branes}
It was shown in \sark\ that the non-maximally symmetric D-branes
introduced in \malmosei\ can be represented as a group product
of the T-dualized U(1) subgroup with a conjugacy class,
\eqn\nbrrep{g=Lhfh^{-1},} 
where $L=e^{-i\alpha\sigma_3}$, $h\in G$ and $f$ is
a constant element.
In this parametrisation,
the topological two-form $\omega^{(2)}$ required to trivialize
the WZW three-form on the brane, 
\eqn\trcon{\omega^{\rm WZ}(g)_{|{\rm brane}}
=d\omega^{(2)},}
can be written as
\eqn\omwznm{\omega^{(2)}(L,h)=\omega^f(h)-{\rm Tr}(L^{-1}dLdCC^{-1}),}
where $\omega^f(h)=h^{-1}dhfh^{-1}dhf^{-1}$.
Here we explore this representation for branes on
the SU(2) group manifold.
In particular we are interested in checking if this
representation is one-to-one.
Given an element $g$, $L$ can be found from the condition 
\eqn\cond{{\rm Tr}(L^{-1}g)={\rm Tr}f.}
Parameterizing g as
\eqn\grparam{g=\left(\matrix{
x_0+ix_3&ix_1-x_2\cr
ix_1+x_2&x_0-ix_3\cr}\right),}
where $x_0^2+x_1^2+x_2^2+x_3^2=1$,
and taking $f$ as $f=e^{i\hat{\psi}\sigma_3}$
we obtain from \cond\
\eqn\condpar{x_0\cos\alpha-x_3\sin\alpha=\cos\hat{\psi}.}
Using now the angles $\theta$, $\phi$ and $\tilde{\phi}$,

\eqn\euleranggg{\eqalign{
& x_0=\cos\theta\cos\tilde{\phi}\cr
& x_1=\sin\theta\cos\phi\cr
& x_2=\sin\theta\sin\phi\cr
& x_3=\cos\theta\sin\tilde{\phi},}}
we can write \condpar\ as 
\eqn\condpel{\cos\theta\cos(\alpha+\tilde{\phi})=\cos\hat{\psi}.}
We see that, given the condition $\cos^2\theta\geq\cos^2\hat{\psi}$
satisfied by the element of the brane \malmosei\ ,
$\alpha$ generically ($\cos\hat{\psi}\neq 0$) is double-valued and equals
\eqn\alphdet{\alpha_{1,2}=
\pm\arccos\left({\cos\hat{\psi}\over \cos\theta}\right)-\tilde{\phi}.}
Consequently every element of the brane can be written in two ways,
\eqn\doubva{g=L_1C_1=L_2C_2,}
where 
\eqn\usubgr{L_{1,2}=e^{-i\alpha_{1,2}\sigma_3}=
\left(\matrix{
{\cos\hat{\psi}\mp iA
\over \cos\theta} e^{i\tilde{\phi}} & 0\cr
0 & {\cos\hat{\psi}\pm iA
\over \cos\theta} e^{-i\tilde{\phi}}\cr}\right),}

\eqn\conjcl{C_{1,2}=
\left(\matrix{
\cos\hat{\psi}\pm iA &
i\tan\theta(\cos\hat{\psi}\pm
iA)e^{-i(\tilde{\phi}-\phi)}\cr
i\tan\theta(\cos\hat{\psi}\mp
iA)e^{i(\tilde{\phi}-\phi)}&
\cos\hat{\psi}\mp iA\cr}
\right)}
and $A=\sqrt{\cos^2\theta-\cos^2\hat{\psi}}$.

It follows from this analysis that generic branes are topologically  
$(U(1)\times S^2)/Z_2$, where $Z_2$ is the identification of the pairs
($L_1, C_1$) and ($L_2, C_2$).
As expected, the subset invariant under this
identification $\cos\hat{\psi}=\cos\theta$ is the boundary
of the branes.

For even $k$ values there exist branes, given by
$\hat{\psi}={\pi \over 2}$, covering the whole group.
For these branes 
there exists the dangerous region $\cos\theta=0$, where $\alpha$
is not defined. Looking at eq. \condpar\ and noting
that for the given values of $\theta$ and $\hat{\psi}$
the left and right sides equal zero, we see that in this region
\condpar\ can be satisfied for any value of $\alpha$. 
The existence of such a region for the branes covering the whole group
was actually expected, because we know
that $\omega^{\rm WZ}(g)$ defines a non-trivial cohomology
class and eq. \trcon\ cannot be satisfied everywhere.
We are led to the conclusion that the brane covering the whole group 
is the complement of the one-dimensional circle
defined by the equation $x_0=x_3=0$. It should be mentioned  that this brane
was already found using a different approach in \klsev\ .

\newsec{The DBI action}
In this section we compute the DBI action on the branes
using the formula \boris, \stanold\ :
\eqn\dbiact{S_{\rm DBI}=\int\sqrt{{\rm det}(G+\omega^{(2)})}.}

In the appendix we show that in the coordinates 
\euleranggg\ $\phi$ , $\tilde{\phi}$ and $\theta$
the two-form $\omega^{(2)}$ takes the form

\eqn\omtwo{{\omega^{(2)}\over k\alpha'}=F_{\theta\phi}d\theta\wedge d\phi
+{\cos2\theta-1\over 2} d\phi\wedge d\tilde{\phi},}
where
\eqn\fielstr{F_{\theta\phi}=
\pm{\tan\theta\cos\hat{\psi}\over 
\sqrt{\cos^2\theta-\cos^2\hat{\psi}}}=
\pm{2\tan\theta\cos\hat{\psi}\over 
\sqrt{2(\cos2\theta-\cos2\hat{\psi})}}.}

The plus and minus signs correspond to different signs in 
\alphdet\ .
Now we argue that only the positive sign should be taken.
For this purpose we compute the restriction of the two-form
\omtwo\ to the two-sphere conjugacy class given by the 
condition $x_0=\cos\hat{\psi}$.
This should result in  the well-known topological two-form for 
conjugacy classes
derived in \alschom\ , \gaw\ and \bdsnew\ .
It is useful to introduce coordinates $\gamma$ and $\beta$
on the two-sphere :
\eqn\eulerangg{\eqalign{
& x_0=\cos\hat{\psi}\cr
& x_1=\sin\hat{\psi}\sin\gamma\cos\beta\cr
& x_2=\sin\hat{\psi}\sin\gamma\sin\beta\cr
& x_3=\sin\hat{\psi}\cos\gamma.}}

Using the formulae \euleranggg\
we can express $\theta$, $\phi$, and $\tilde{\phi}$
as functions of $\gamma$ and $\beta$.
Equating $x_0^2+x_3^2$ from \eulerangg\ and  \euleranggg\
we obtain:
\eqn\reform{\cos^2\theta=\cos^2\hat{\psi}+\sin^2\hat{\psi}\cos^2\gamma,}
or
\eqn\reformm{\sin\theta=\sin\hat{\psi}\sin\gamma.}

Equating $x_0$ from \eulerangg\ and \euleranggg\
and substituting \reform\ we obtain:
\eqn\cosre{\cos\tilde{\phi}={\cos\hat{\psi}\over
\sqrt{\cos^2\hat{\psi}+\sin^2\hat{\psi}\cos^2\gamma}}}
and
\eqn\sinre{\sin\tilde{\phi}={\sin\hat{\psi}\cos\gamma\over
\sqrt{\cos^2\hat{\psi}+\sin^2\hat{\psi}\cos^2\gamma}}.}
 
Equating $x_1$ and $x_2$ from \eulerangg\ and \euleranggg\
and substituting 
\reformm\ we note that we can set 
\eqn\lastnew{\phi=\beta}
Using \reform\ ,\reformm\ ,\cosre\ ,\sinre\ and \lastnew\
it is straightforward to compute
\eqn\restsph{F_{\theta\phi}d\theta\wedge d\phi=
\pm{\sin\theta\cos\hat{\psi}\over \cos^2\theta}d\gamma\wedge d\beta}
and 
\eqn\resttt{
{\cos2\theta-1\over 2} d\phi\wedge d\tilde{\phi}=
-{\sin^3\theta\cos\hat{\psi}\over \cos^2\theta}d\gamma\wedge d\beta.}
Putting together \restsph\ , \resttt\ , we finally get
\eqn\twoff{
\omega^{(2)}|_{x_0=\cos\hat{\psi}}={\sin\theta\cos\hat{\psi}\over \cos^2\theta}
(\pm 1-\sin^2\theta)d\gamma\wedge d\beta.}
We see that taking the positive sign in \twoff\ we obtain
\eqn\twforres{
\omega^{(2)}|_{x_0=\cos\hat{\psi}}=k\alpha'\sin\theta\cos\hat{\psi}
d\gamma\wedge d\beta=k\alpha'\sin\gamma\sin\hat{\psi}\cos\hat{\psi}
d\gamma\wedge d\beta
=k\alpha'{\sin\gamma\sin2\hat{\psi}\over 2}d\gamma\wedge d\beta,}
which is exactly the two-form for the conjugacy class found
in \alschom\ , \gaw\ and \bdsnew\ . 

We  note that $F_{\theta\phi}d\theta\wedge d\phi$  coincides with 
the field strength found in \malmosei\ for the B-brane
on the parafermion disk. (One should take into account 
that the radius $\rho$ in that
paper is $\rho=\sin\theta$.)
This can be explained by noting that, as shown in  \walzh\ , the B-branes 
on the parafermion disk can be derived by the axial
gauging of the B-branes on group manifolds. Fixing the gauge
 $\tilde{\phi}=0$  and using Stokes' theorem 
we obtain the boundary term for the B-brane
on the parafermion disk from the boundary term of the  WZW model.

We now turn to the computation of the DBI action and mass of
the branes. 

In the coordinates $\phi$ , $\tilde{\phi}$
and $\theta$ the metric is 
\eqn\metric{ ds^2 = k\alpha^\prime (d\theta^2 + {\rm cos}^2\theta\; 
d\tilde{\phi}^2 +
{\rm sin}^2\theta\; d\phi^2)\ ,} 

For $\omega^{(2)}$ given by \omtwo\ and the metric given by \metric\ ,
the $G+\omega^{(2)}$ matrix is
\eqn\matr{
G+\omega^{(2)}=k\alpha^\prime\left(\matrix{
1 &F_{\theta\phi} & 0\cr
-F_{\theta\phi}& {\rm sin}^2\theta & -{\rm sin}^2\theta\cr
0 & {\rm sin}^2\theta & {\rm cos}^2\theta\cr}\right).}

Using \fielstr\ we  compute that 
\eqn\endb{
\sqrt{{\rm det}(G+\omega^{(2)})}=(k\alpha')^{3/2}
\sqrt{\sin^2\theta+F_{\theta\phi}^2
\cos^2\theta}=(k\alpha')^{3/2}{\sin2\theta\over 
\sqrt{2(\cos2\theta-\cos2\hat{\psi})}}.}
This result is in agreement with the computation of the overlap
of the boundary state with the graviton wave packet in \malmosei\ :
\eqn\loc{\langle B,j,\eta|\theta\rangle\sim
{\Theta(\cos2\theta-\cos2\hat{\psi})\over
\sqrt{\cos2\theta-\cos2\hat{\psi}}}.}

For the energy we obtain
\eqn\enrgy{
E_{\rm DBI}=T_{(3)}\int_0^{\hat{\psi}}d\theta\int_0^{2\pi}d\phi\int_0^{2\pi}
d\tilde{\phi}
\sqrt{{\rm det}(G+\omega^{(2)})}=4\pi^2(k\alpha')^{3/2}T_{(3)}\sin\hat{\psi}.}
Recalling now that the masses  of A-branes wrapping 
conjugacy classes are equal to
\eqn\massabr{
M(A,\hat{\psi})=4\pi k\alpha' T_{(2)}\sin\hat{\psi}}
and the relation between the tensions of the D2- and D3-branes, 
\eqn\ten{\pi\sqrt{\alpha'}T_{(3)}=T_{(2)},}
we get
\eqn\massab{M(B,\hat{\psi})=\sqrt{k}M(A,\hat{\psi})}
as predicted by the CFT computation.

\vskip 30pt

{\bf Acknowledgement} 

\vskip 10pt

The author thanks C. Bachas, M. Blau, K. S. Narain and C. Schweigert 
for useful discussions.

\vfill

\appendix{A}{The $\omega^{(2)}$ computation}
Here we use the Euler angles $\chi$ , $\varphi$ and $\tilde{\theta}$ 
connected to the coordinates \euleranggg\ by the formulae
\eqn\newcor{\eqalign{
&\chi=\tilde{\phi}+\phi\cr
& \varphi=\tilde{\phi}-\phi\cr
& \tilde{\theta}=2\theta.}}
In terms of Euler angles,  $g$ has the form
\eqn\euang{
g=\left(\matrix{
\cos{\tilde{\theta}\over 2}e^{i{\chi+\varphi\over 2}}&
i\sin{\tilde{\theta}\over 2}e^{i{\chi-\varphi\over 2}}\cr
i\sin{\tilde{\theta}\over 2}e^{-i{\chi-\varphi\over 2}}&
\cos{\tilde{\theta}\over 2}e^{-i{\chi+\varphi\over 2}}}\right).}
Parametrizing $g$ as $g=LC=Lhfh^{-1}$ , where $f=e^{i\hat{\psi}\sigma_3}$ , we have
\eqn\wzonbound{
\omega^{(2)}(L,h)=\omega^f(h)-{\rm Tr}(L^{-1}dLdCC^{-1}).}

Using \usubgr\ and \conjcl\ we compute
\eqn\ualg{
L_{1,2}^{-1}dL_{1,2}=\left(\matrix{
\pm{i\cos\hat{\psi}\tan{\tilde{\theta}\over 2}\over 2A}d\tilde{\theta}+
i{d\chi+d\varphi\over 2} & 0\cr
0 & \mp i{\cos\hat{\psi}\tan{\tilde{\theta}\over 2}\over 2A}d\tilde{\theta}
-i{d\chi+d\varphi\over 2}\cr}\right),}

\eqn\algconj{
dC_{1,2}C_{1,2}^{-1}=\left(\matrix{
\mp i{\cos\hat{\psi}\tan{\tilde{\theta}\over 2}\over 2A}d\tilde{\theta}
-i\sin^2{\tilde{\theta}\over 2}d\varphi& *** \cr
*** & 
\pm i{\cos\hat{\psi}\tan{\tilde{\theta}\over 2}\over 2A}d\tilde{\theta}
+i\sin^2{\tilde{\theta}\over 2}d\varphi\cr}\right).}

The terms denoted by stars are not important for our purposes.

Collecting \ualg\ and \algconj\ we obtain

\eqn\trace{
{\rm Tr}(L_{1,2}^{-1}dL_{1,2}dC_{1,2}C_{1,2}^{-1})=
\pm{\cos\hat{\psi}\tan{\tilde{\theta}\over 2}\over 2A}d\tilde{\theta}
\wedge d\chi\mp{\cos\hat{\psi}\tan{\tilde{\theta}\over 2}\cos\tilde{\theta}
\over 2A}d\tilde{\theta}
\wedge d\varphi+\sin^2{\tilde{\theta}\over 2}d\chi\wedge d\varphi.}

Now we turn to the computation of $\omega^f(h)$.

Let us take $C$ in the form:
\eqn\cparam{
C=\left(\matrix{
c_{11}& c_{12}\cr
c_{21}& c_{22}\cr}\right),}
where
\eqn\cspec{\eqalign{
& c_{11}=\cos\hat{\psi}+ix_3\cr
& c_{12}=ix_1-x_2\cr
& c_{21}=ix_1+x_2\cr
& c_{22}=\cos\hat{\psi}-ix_3,}}
and $x_1^2+x_2^2+x_3^2=\sin^2\hat{\psi}$.
Now we should find an $h$ satisfying the relation $C=hfh^{-1}$,
where $f=e^{i\hat{\psi}\sigma_3}$.
It is easy to check that it is possible to choose an $h$ of the form
\eqn\hparm{h=\left(\matrix{
a & b\cr
-b^{*} & a^{*} \cr}\right),}
where
\eqn\abform{\eqalign{
& a=-{x_1+ix_2\over \sqrt{2\sin\hat{\psi}(\sin\hat{\psi}-x_3)}}\cr
& b=\sqrt{\sin\hat{\psi}-x_3\over 2\sin\hat{\psi}}.}}
Substituting \hparm\ in $\omega^f(h)$ we obtain

\eqn\omabparm{
\omega^f(h)=2i\sin2\hat{\psi}(a^{*}b^{*}da\wedge db-bb^{*}da\wedge da^{*}
-aa^{*}db^{*}\wedge db+abdb^{*}\wedge da^{*}).}
Substituting \abform\ in \omabparm\ we get:
\eqn\omforcc{
\omega^f(h)={dc_{12}\wedge dc_{21}\over ix_3}\cos\hat{\psi}.}

Finally, using \conjcl\ we get
\eqn\omfinal{
\omega^f(h)=\mp{\tan{\tilde{\theta}\over 2}\cos^2{\tilde{\theta}\over 2}
\cos\hat{\psi}
\over \sqrt{\cos^2{\tilde{\theta}\over 2}-\cos^2\hat{\psi}}}d\tilde{\theta}
\wedge d\varphi.}
Inserting now \trace\ and \omfinal\ in \wzonbound\ we obtain:
\eqn\ontwfin{
 \omega^2(L,h)=\pm{\cos\hat{\psi}\tan{\tilde{\theta}\over 2}\over 2
\sqrt{\cos^2{\tilde{\theta}\over 2}\mp\cos^2\hat{\psi}}}d\tilde{\theta}
\wedge d\chi\mp{\cos\hat{\psi}\tan{\tilde{\theta}\over 2}\over 2
\sqrt{\cos^2{\tilde{\theta}\over 2}-\cos^2\hat{\psi}}}d\tilde{\theta}
\wedge d\varphi-\sin^2{\tilde{\theta}\over 2}d\chi\wedge d\varphi.}

Then, using the coordinates \euleranggg\
we can rewrite \ontwfin\ compactly in the form
\eqn\omtwnewcor{
{\omega^2(L,h)\over 2}=\pm{2\cos\hat{\psi}\tan\theta\over 
\sqrt{2(\cos2\theta-\cos2\hat{\psi})}}d\theta
\wedge d\phi-\sin^2\theta d\phi\wedge d\tilde{\phi}.}

It follows from conformal invariance that in the DBI action we should use half
of $\omega^2(L,h)$ as in \omtwnewcor\ .

\listrefs
\end